\documentclass[rapids]{jfm}

\usepackage{graphicx}
\usepackage{epstopdf,epsfig}
\usepackage{newtxtext}
\usepackage{newtxmath}
\usepackage{tikz}
\usetikzlibrary{shapes}
\usepackage{color}
\usepackage{natbib}
\usepackage{hyperref}
\usepackage{tabu}
\usepackage{longtable}
\usepackage[noabbrev]{cleveref}
\usepackage{ragged2e}

\newcommand{\RomanNumeralCaps}[1]
\linenumbers

\usepackage{bm}
\usepackage[final]{changes}
\usepackage{subcaption}

\newcommand{\ub}{{\bf u}}

\newcommand{\Xb}{{\bf X}}

\newcommand{\xb}{{\bf x}}

\newcommand{\St}{\mathit{St}}

\newcommand{\rhoPar}{\rho_\mathrm{s}}
\newcommand{\rhoFib}{\widetilde{\rho}_\mathrm{s}}

\begin{document}

\title{The effect of particle anisotropy on the modulation of turbulent flows}

\author{Stefano Olivieri\aff{1,2},
Ianto Cannon\aff{1}
\and Marco E. Rosti\aff{1}  \corresp{\email{marco.rosti@oist.jp}}}

\affiliation{
\aff{1}Complex Fluids and Flows Unit, Okinawa Institute of Science and Technology Graduate University, 1919-1 Tancha, Onna-son, Okinawa 904-0495, Japan\\
\aff{2}Department of Aerospace Engineering, Universidad Carlos III de Madrid, Avda. de la Universidad, 30. 28911 Legan\'es, Spain
}

\date{\today}

\maketitle

\begin{abstract}
We investigate the modulation of turbulence caused by the presence of finite-size dispersed particles. Bluff (isotropic) spheres vs slender (anisotropic) fibers are considered to understand the influence of the object shape on altering the carrier flow. While at a fixed mass fraction -- but different Stokes number -- both objects provide a similar bulk effect characterized by a large-scale energy depletion, a scale-by-scale analysis of the energy transfer reveals that the alteration of the whole spectrum is intrinsically different. For bluff objects, the classical energy cascade is shrinked in its extension but unaltered in the energy content and its typical features, while for slender ones we find an alternative energy \added{flux} which is essentially mediated by the fluid-solid coupling.
\end{abstract}


\section{Introduction}

Particle-laden turbulent flows are multiphase systems where a carrier fluid interacts with a dispersed phase made by a number of solid objects, e.g., spheres or fibers. Such flows concern an important class of problems with numerous applications related to both natural and industrial processes~\citep{delillo2014turbulent,breard2016coupling,sengupta2017phytoplankton, falkinhoff2020preferential, rosti2020fluid}.
In the analysis and modelling of such problems, a crucial distinction can be made regarding the mutual coupling between the carrier flow and the dispersed objects. When the suspension is dilute enough, it can be safely assumed that the fluid flow is not substantially altered by the presence of the objects~\citep{balachandar2010turbulent,maxey2017simulation,brandt2021particle}. However, we often deal with non-dilute conditions where the mutual coupling between the two phases is relevant and gives rise to a macroscopic alteration of the turbulent carrier flow. The resulting turbulence modulation effects have been the subject of previous studies over different classes of multiphase turbulent flows, i.e., considering isotropic~\added{\citep{lucci2010modulation,gualtieri2013clustering,uhlmann_clustering_2017,capecelatro2018transition, ardekani2019turbulent,yousefi_modulation_2020}} or anisotropic~\citep{andersson2012torque,olivieri2020dispersed,olivieri2020turbulence, olivieri2021universal, olivieri2022fully, wang2022finite} solid particles, as well as droplets or bubbles~\added{\citep{dodd2016interaction,rosti2019droplets,freund2019wavelet,cannon2021effect}}, typically focusing on the alteration of both the bulk flow properties as well as the scale-by-scale energy distribution. \added{In particular, \cite{lucci2010modulation} and \cite{yousefi_modulation_2020} showed that Taylor length-scale sized spheres reduce turbulent kinetic energy at the large scales and enhance its energy content at the small scales.} Nevertheless, the accurate characterization of the underlying physics in these complex systems still requires significant efforts both from the theoretical, computational and experimental viewpoint with relevant questions still not fully addressed, such as: (i) What are the mechanisms controlling the scale-by-scale energy distribution in the presence of immersed objects with finite size (i.e., larger than the dissipative lengthscale)? (ii) How do the geometrical properties of the dispersed particles (i.e., their size and isotropy) affect the backreaction on the carrier flow and the consequent turbulence modulation?

\begin{figure}
\centering
\includegraphics[width=0.6\textwidth]{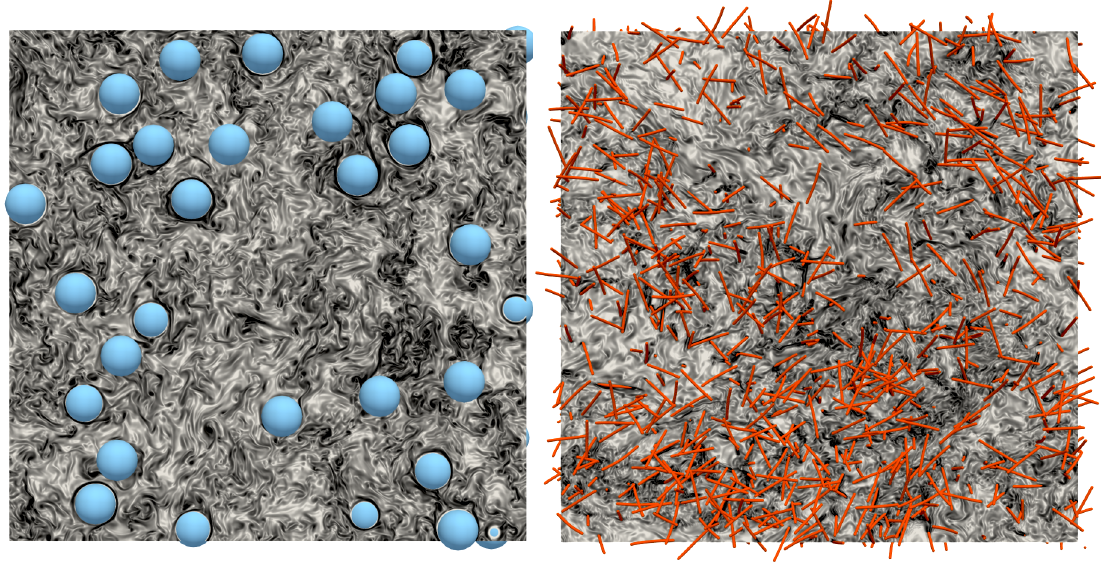}
\caption{
 2-D views of the vorticity magnitude of homogeneous isotropic turbulence in the presence of dispersed, finite-size spheres (left) and fibers (right), from two representative cases of the present DNS study.
 }
 \label{fig:snapshots}
\end{figure}

In this work, \added{we comprehensively investigate the multiscale nature of the turbulence modulation due to finite-size rigid particles, focusing on the role of geometrical properties and comparing, in particular, the backreaction caused by isotropic bluff objects (i.e., spheres) vs anisotropic slender ones (i.e., fibers)}. Exploiting massive direct numerical simulations (DNS) it is observed, at first, that the macroscopic effect in the turbulence modulation essentially consists of a large-scale energy depletion for both configurations. However, we show that this bulk effect arises from qualitatively different mechanisms depending on the geometrical features of the dispersed objects, which becomes evident from a scale-by-scale energy-transfer balance. For isotropic objects (spheres), the backreaction effectively acts at a well defined lengthscale (i.e., the sphere diameter) and over a limited range of smaller scales, without appreciably modifying the inertial range that is obtained in the single phase (i.e., without particles) configuration. For anisotropic objects (fibers), instead, the fluid-solid coupling is responsible for a global modification of the energy distribution over all the scales of motion, which is characterized by the emergence of an alternative energy \added{flux} along with a relative enhancement of small-scale fluctuations.

The rest of the paper is structured as follows: \cref{sec:meth} describes the modelling and computational methodology, \cref{sec:resu} shows the results and \cref{sec:conc} contains the conclusions.

\section{Methods}
\label{sec:meth}
  
\begin{figure}
    \centering
    \includegraphics[width=1.0\textwidth]{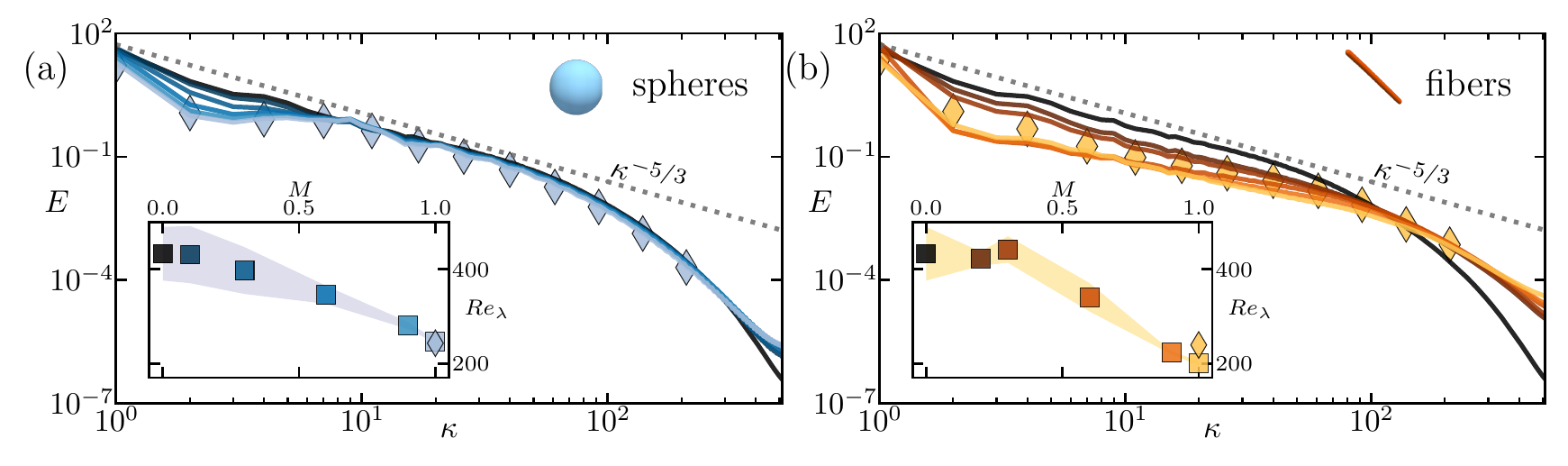}
    \caption{
   Energy spectra of the modulated turbulent flow for (\emph{a}) spheres and (\emph{b}) fibers, for different mass fraction $M$ (increasing with the color brightness from dark to light), along with the reference single phase configuration (i.e., $M=0$, black curve) and the expected Kolmogorov scaling in the inertial subrange (gray dashed line). 
   The insets report the micro-scale Reynolds number $\Rey_\lambda$ as a function of the mass fraction; error bars show the standard deviation in $\Rey_\lambda$ from the time averaged value.
   As an additional check on the accuracy of the computations, diamonds show results calculated using an Eulerian grid with halved resolution ($512^3$ cells), which produces little change in $\Rey_\lambda$ and the inertial range of the spectra.
  }
    \label{fig:spectra}
\end{figure}

To investigate the problem, we devote our attention to particles of finite size (i.e., diameter or length) which lies well within the inertial subrange of the turbulent flow. A visual example of two representative configurations is given in~\cref{fig:snapshots}.
Specifically, we have performed DNS where the fluid and solid dynamics are mutually coupled using the immersed boundary method~\citep{olivieri2022fully,hori2022eulerian}.
An incompressible, homogeneous and isotropic turbulent (HIT) flow is generated within a triperiodic cubic domain of size $L=2\pi$ using the Arnold-Beltrami-Childress (ABC) cellular-flow forcing~\citep{podvigina1994non}, achieving in the single phase case a micro-scale Reynolds number $\Rey_\lambda = u' \lambda / \nu \approx 435$, where $u'$ is the root mean square of the turbulent fluctuations, $\lambda$ the Taylor's micro-scale, and $\nu$ the kinematic viscosity. 
\added{Such high-Reynolds-number configuration is computationally explored for the first time in the framework of multiphase flows in order to achieve proper scale separation.} As shown in \cref{fig:spectra}, the energy spectrum in the single phase configuration (black curve) shows the classical Kolmogorov scaling $\sim \kappa^{-5/3}$ (dashed line) at low-to-intermediate wavenumbers over more than one decade.

\begin{table}
\centering

  \begin{tabular}{lccccccc|lccccccc}
    $\rho_\mathrm{s} / \rho_\mathrm{f}$ & $D/L$ & \added{$D/\eta$} & $\St$ & $N$ & $M$ & $\Rey_\lambda$ & $C_d$ &    $\Delta \widetilde{\rho} / (\rho_\mathrm{f} L^2)$ & $c/L$ & \added{$c/\eta$} & $\St$ & $N$ & $M$ & $\Rey_\lambda$ & $C_d$ \\[3pt]
    $1.3$ & $(4\pi)^{-1}$ & \added{$123$} &7.4 & $300$ & $0.1$ & $431$ & $0.11$    & $3.0\times10^{-4}$ & $(4\pi)^{-1}$ & \added{$123$} &0.45 & $10^4$ & $0.2$ & $422$ & 0.11 \\
    $5.0$ & $(4\pi)^{-1}$ & \added{$123$} &27 & $300$ & $0.3$ & $397$ & $0.13$    & $5.3\times10^{-4}$ & $(4\pi)^{-1}$ & \added{$123$} &0.79 & $10^4$ & $0.3$ & $442$ & 0.11\\
    $1.7\times10^{1}$ & $(4\pi)^{-1}$ & \added{$123$} &90 & $300$ & $0.6$ & $346$ & $0.16$  &   $1.9\times10^{-3}$ & $(4\pi)^{-1}$ & \added{$123$} &2.8 & $10^4$ & $0.6$ & $340$ & 0.17 \\
    $1.0\times10^{2}$ & $(4\pi)^{-1}$ & \added{$123$} &470 &  $300$ & $0.9$ & $280$ & $0.21$ &     $1.1\times10^{-2}$ & $(4\pi)^{-1}$ & \added{$123$} &16 & $10^4$ & $0.9$ & $223$ & 0.29 \\
    $\infty$ & $(4\pi)^{-1}$ & \added{$123$} &$\infty$ & $300$ & $1$ & $247$ & $0.23$ & $\infty$ &                 $(4\pi)^{-1}$ & \added{$123$} &$\infty$ & $10^4$ & $1$ & $200$ & 0.32 \\
  \end{tabular}\\[9pt]

\caption{Parametric combinations investigated in our baseline study. Left: suspensions of bluff, spherical particles ($\rho_\mathrm{s}$ is the volumetric density of solid phase, $D$ the sphere diameter). Right: suspensions of anisotropic, slender particles ($\Delta \widetilde{\rho}$ is the linear density difference between the solid and fluid phases, $c$ the fiber length). Here, \added{$\eta$ is the Kolmogorov microscale of the single phase case,} $\St$ is the estimated Stokes number of the particle, $N$ the number of dispersed particles, $M$ the corresponding mass fraction, $\Rey_\lambda$ and $C_d$ the resulting micro-scale Reynolds number and drag coefficient of the modulated flow, respectively (for $M=0$, $\Rey_\lambda \approx 435$ and $C_d \approx 0.12$). The cases with $\rho_\mathrm{s}=\infty$ and $\Delta \widetilde{\rho}=\infty$ correspond to the configurations where the particles are retained fixed.
 In addition to the cases reported in the table, we have performed another set of simulations in the fixed-particle arrangement varying $N$ and $D$ or $c$ (the results of which are reported in~\cref{fig:source}).}
\label{tab:parCases}
\end{table}

Once the single phase case reached the fully-developed regime, $N$ rigid spheres (characterized by diameter $D$ and volumetric density $\rhoPar$) or fibers (characterized by length $c$ and linear density difference $\Delta \rhoFib$) are added to the carrier flow at randomly initialised positions and orientations. The multiphase cases were therefore evolved until reaching a statistically stationary state. An overview of the main configurations considered in our study is shown in \cref{tab:parCases}. The dynamics of the bluff, spherical objects is governed by the well-known Euler-Newton equations~\citep{hori2022eulerian}, whereas the slender, anisotropic ones are modelled in the general framework of the Euler-Bernoulli equation for inextensible filaments, choosing a sufficiently large bending stiffness such that the deformation is always negligible (i.e., within 1\%)~\citep{cavaiola2019assembly,brizzolara2021fiber}.
Hence, the mass fraction $M$ of the suspension is defined as the ratio between the mass of the dispersed solid phase and the total mass (i.e., the sum of the fluid and solid mass contained in the domain).
Note that for the chosen parameters and when matching $M$, bluff and slender particles have remarkably different Stokes number, here computed using expressions for small particles, i.e., of length below the dissipative scale, for the sake of a comparative estimate~\citep{lucci2011stokes,bounoua2018tumbling}. 
Moreover, we indicate $M=1$ as the configurations where the dispersed objects are constrained to a fixed random position; such setting serves as the limiting case where the dispersed phase has infinitely large inertia, as well as a representation of flows in porous media.
Finally, we note that sphere and fiber cases with same mass fraction have different volume fraction but approximately the same total wetted area. 

To solve the governing equations numerically, we employ the in-house solver \textit{Fujin} (\texttt{https://groups.oist.jp/cffu/code}).
The code is based on the (second-order) central finite-difference method for the spatial discretization and the (second-order) Adams-Bashforth scheme for the temporal discretization. The incompressible Navier-Stokes equations are solved using the fractional step method on a staggered grid. The Poisson equation enforcing the incompressibility constraint is solved using a fast and efficient approach based on the Fast Fourier Transform (FFT).
The solver is parallelized using the MPI protocol and the \texttt{2decomp} library for domain decomposition (\texttt{http://www.2decomp.org}). 
In this work, the fluid domain is discretized onto a uniform Eulerian grid using $1024^3$ cells, ensuring that, for the chosen set of domain size and fluid properties, the ratio between the Kolmogorov dissipative lengthscale and the grid spacing is $\eta/\Delta x = \mathcal{O}(1)$.
The carrier- and dispersed-phase dynamics are coupled by the no-slip condition $\dot{\Xb} = \mathbf{U} = \ub \left( \Xb,t \right)$, where $\Xb$ is the position of a generic material point on the solid surface and $\ub=\ub\left( \xb,t \right)$ is the fluid velocity field. In the present work, we employ two types of immersed boundary (IB) method where the mutual interaction between the two phases is achieved by means of a singular force distribution. Specifically, for bluff spherical particles we use the Eulerian IB method recently proposed by~\citet{hori2022eulerian}, whereas for slender fibers we use the method originally proposed by~\citet{huang_shin_sung_2007a} and recently employed for fiber-laden turbulence by~\citet{olivieri2020dispersed,olivieri2020turbulence,olivieri2021universal,olivieri2022fully}.
Overall, the code has been extensively validated and tested in a variety of problems, see e.g.~\cite{rosti2020increase,rosti2020fluid,rosti2021turbulence,olivieri2022fully}.

\section{Results}
\label{sec:resu}

\subsection{Main features of turbulence modulation}

The presence of the dispersed phase clearly causes a complex modification of the key features of the carrier flow, as it can be observed in the energy spectra for suspensions of spheres (\cref{fig:spectra}\textit{a}) or fibers (\cref{fig:spectra}\textit{b}) at different mass fraction $M$.
At a first glance and focusing on the smallest wavenumbers (i.e., largest scales), one can note a similar phenomenology between the two kinds of particles, with an overall tendency to decrease the turbulent kinetic energy while increasing $M$. Indeed, for both bluff and slender particles the energy-containing scales are depleted by the hydrodynamic drag exerted by the particles. 
A direct indication on how the bulk properties of the flow are altered is provided in the insets of~\cref{fig:spectra}, showing a very similar variation between spheres and fibers in terms of $\Rey_\lambda$ with the mass fraction, notwithstanding the different Stokes numbers of the suspended objects, in agreement with previous findings~\citep{hwang2006homogeneous,olivieri2021universal,olivieri2022fully}. 

However, from~\cref{fig:spectra} some peculiar differences between the two kinds of suspensions can also be noticed when extending the observation to the full range of active scales. For bluff particles, the alteration of the energy spectrum with $M$ remains \added{almost entirely} limited to the low-wavenumber region (i.e., $\kappa \lesssim 5$)\added{, with only a minimal increase at the largest wavenumbers (i.e., $\kappa \gtrsim 300$) associated to the high shear regions in the boundary layers around the spheres. Instead,}  for fibers the modulation extends up to the highest wavenumber (i.e., $\kappa_\mathrm{max} = 512$). At sufficiently large $M$, a departure from the Kolmogorov scaling indeed appears throughout the full inertial subrange. We anticipate that here the energy transfer \added{is mainly} due to the fluid-solid coupling, and not to the convective term as in the single phase or bluff particle cases, \added{leading to a different form of energy flux}. 

 \begin{figure}
    \centering
    \includegraphics[width=.6\textwidth]{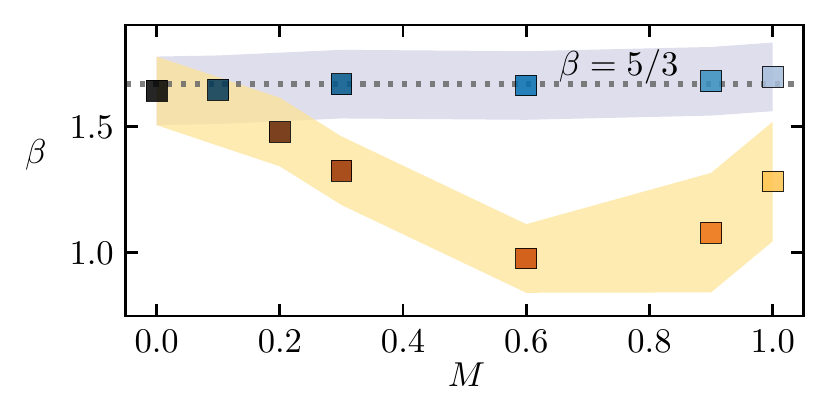}
    \caption{
Dependence of the exponent $\beta$ in the energy spectrum scaling $E\sim \kappa^{-\beta}$ on particle mass fraction $M$. Flows with spheres are marked in blue, flows with fibers in orange, and the single phase flow in black. The blue and orange shaded regions show the approximate error in $\beta$, estimated by moving the time averaging window. The Kolmogorov scaling is marked by a grey dotted line.}
    \label{fig:scalExp}
\end{figure}

A quantitative evaluation of the resulting power law $E(\kappa) \sim \kappa^{-\beta}$ in the inertial subrange of both single and multiphase flows is given in \cref{fig:scalExp}, showing the scaling exponent $\beta$ as a function of the mass fraction for both spheres and fibers.
The single phase flow ($M=0$) and the flows with spheres can be seen to follow the Kolmogorov scaling ($\beta=5/3$), whereas the flows with fibers show a significant reduction in $\beta$ as $M$ increases. 
An heuristic explanation for the latter trend is that fibers act as a barrier to the flow between any two points with separation greater than the fibre diameter $d$, which influences the scaling of the second order velocity structure function $\langle (\delta u)^2 \rangle \sim r^\gamma$ for two points at a distance $r > d$, with $\gamma = \beta - 1$. In the single phase case $\gamma = 2/3$, whereas the presence of fibers tend to decorrelate the flow, thus reducing the value of $\gamma$ or, equivalently, $\beta$.

\begin{figure}
    \centering
    \includegraphics[width=0.9\textwidth]{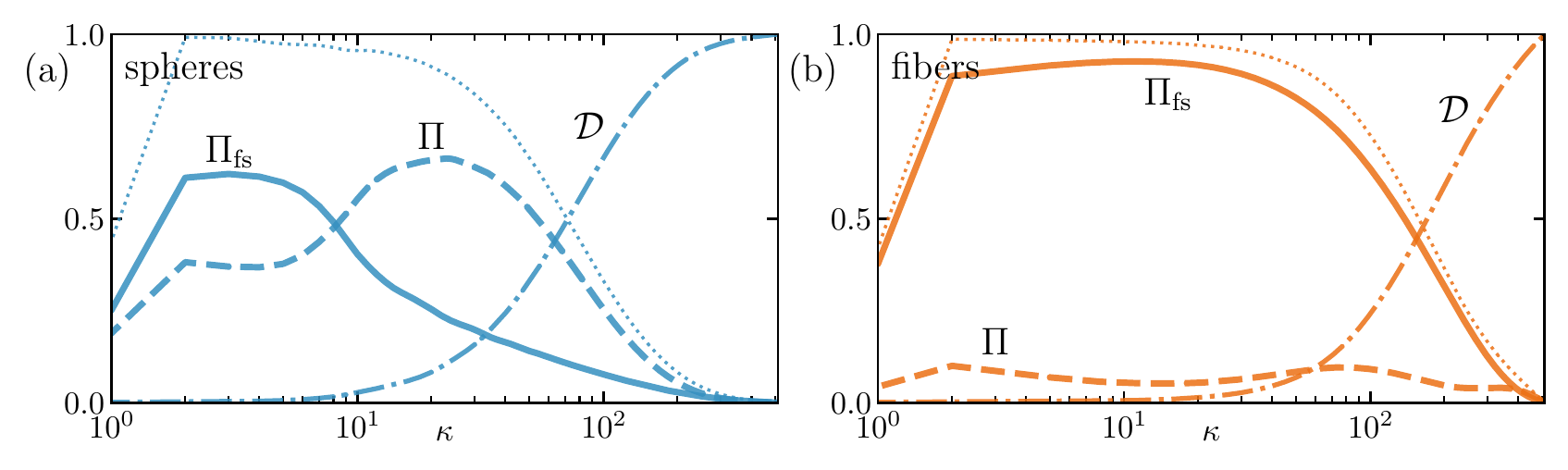}
    \caption{
    Scale-by-scale energy transfer balance for two representative configurations at $M=0.9$ of (\emph{a}) spheres and (\emph{b}) fibers, showing the fluid-solid coupling $\Pi_\mathrm{fs}$ (solid line), nonlinear convective $\Pi$ (dashed line) and viscous dissipation $\mathcal{D}$ (dash-dotted line) contributions, each normalized with the average dissipation rate $\epsilon$. Furthermore, the total energy flux, $\Pi_\mathrm{fs} + \Pi$, is also reported (dotted line).
    }
    \label{fig:fluxes}
\end{figure}

\subsection{Scale-by-scale energy transfer}

A clear distinction in the mechanism of energy distribution between the two geometrical configurations can be highlighted and
to gain a more detailed insight we look at the scale-by-scale energy transfer balance
\begin{equation}
  \mathcal{P}(\kappa) + \Pi (\kappa) + \Pi_\mathrm{fs}(\kappa) + \mathcal{D}(\kappa) = \epsilon,
  \label{eq:energy-transfer}
\end{equation}
where $\mathcal{P}$ is the turbulence production associated with the external forcing (acting only at the largest scale $\kappa=1$), $\Pi$ and $\Pi_\mathrm{fs}$ are the energy fluxes associated with the nonlinear convective term and the fluid-solid coupling term, respectively, and $\mathcal{D}$ is the viscous dissipation~\citep{olivieri2021universal,olivieri2022fully}. In~\cref{fig:fluxes}, we show the energy fluxes and dissipation in two representative cases with strong backreaction ($M=0.9$) for (\emph{a}) bluff and (\emph{b}) slender particles. 
Focusing on the two different energy fluxes (i.e., $\Pi$ and $\Pi_\mathrm{fs}$), we note at first that the sum of these two contributions (thin dotted line) appears in both cases as a horizontal plateau for relatively low wavenumbers, as expected from~\cref{eq:energy-transfer} and similar to the single phase case. However, qualitatively different scenarios can be identified for bluff vs slender objects when analyzing separately the two distinct contributions. On the other hand, it can be noticed that in both cases, and similarly to the classical, single phase case, for sufficiently large wavenumbers the energy fluxes tend to zero, and the viscous dissipation $\mathcal{D}$ recovers the totality of the balance.
 
For bluff objects (\cref{fig:fluxes}\emph{a}), we first have a dominance of the fluid-solid coupling $\Pi_\mathrm{fs}$  contribution within a limited low-wavenumber range, and only subsequently of the convective term $\Pi$  for larger $\kappa$. Indeed, two distinct plateau-like regions are found over two distinct subranges of scales, suggesting that, for increasing $\kappa$, the energy is first transferred from the largest scales (where energy is injected) to smaller ones mainly by the action of the particles, only after which the nonlinear term prevails and the balance substantially recovers the classical energy cascade predicted by Kolmogorov theory. For slender objects (\cref{fig:fluxes}\emph{b}), the scenario looks radically different, with $\Pi_\mathrm{fs}$ acting over a much wider range of scales and being responsible for transferring most of the energy across all scales, with the nonlinear term being overall weakened. It can also be noted that such \added{alternative energy flux} is overall prolonged with respect to the single phase case,
consistently with the observed alteration in the energy spectrum (\cref{fig:spectra}\emph{b}). Note that we refer to an energy \added{flux} also for fiber-laden turbulence because not only \added{$\Pi_\mathrm{fs}$} is constant across a wide range of scales, but also the overall drag coefficient $C_d = \epsilon / (u'^3 \kappa_\mathrm{in})$, where $\kappa_\mathrm{in}=1$ is the wavenumber at which the energy is injected~\citep{alexakis2018cascades}, remains finite and comparable with the single phase case (see~\cref{tab:parCases}). 

\subsection{Characteristic lengthscale of the fluid-solid coupling}
 
\begin{figure}
    \centering
    \includegraphics[width=\textwidth]{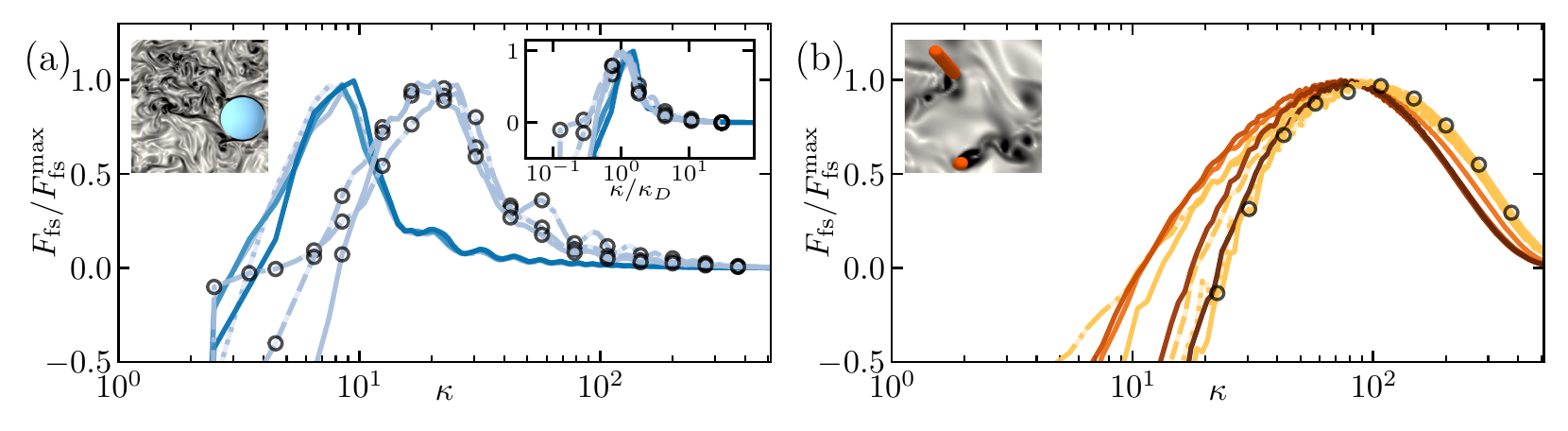}
    \caption{
    Fluid-solid coupling contribution to the energy-spectrum balance for (\emph{a}) spheres and (\emph{b}) fibers for various mass fractions $M$ (varying with color brightness). Circles are plotted for the cases with (\emph{a}) smaller diameter $D$ or (\emph{b}) shorter length $c$, while the different line style is used to denote the variation of the number of objects $N$. For ease of comparison, the $y$-axis is normalized by the maximum value of the reported quantity. The inset in (\emph{a}) shows the same data as a function of the wavenumber $\kappa$ normalized with the sphere diameter $D$. Images show wakes which are similar in size to the (\emph{a}) sphere diameter, and (\emph{b}) fiber diameter.
    }
    \label{fig:source}
\end{figure}

The reason for the observed difference between bluff and slender objects can be ascribed indeed to specific geometrical features. For bluff, isotropic particles the most representative scale is uniquely identified as the particle diameter $D$. For slender fibers, the backreaction could be expected instead to act across multiple lengthscales, approximately ranging from the fiber length $c$ to the cross-sectional diameter $d$. In fact, the latter is found to have the dominant role (as later shown in \cref{fig:source}). This qualitative difference has a remarkable consequence on the  properties of the modulated turbulent flow at sufficiently small scales: on one hand, spherical particles affect the flow essentially only at a scale that is well within the inertial range, without modifying the extension and energy amplitude of the latter; on the other hand, slender fibers are directly acting on wavenumbers that are also beyond the original energy cascade.

To isolate the characteristic lengthscale up to which the energy is transferred by the backreaction for the two kinds of dispersed objects, we show in~\cref{fig:source} the fluid-solid coupling contribution in the energy-spectrum balance, i.e., $F_\mathrm{fs}$ such that $\int_\kappa^\infty  F_\mathrm{fs} = \Pi_\mathrm{fs}$. To this aim, along with the variation of the mass fraction, we also consider the influence of the sphere diameter $D$ or fiber length $c$ in the limiting case of fixed objects (or infinite inertia). For bluff (isotropic) objects (\cref{fig:source}\emph{a}), it can be clearly observed that the peak of $F_\mathrm{fs}$ scales with the diameter $D$, as also shown from the panel inset where the wavenumber is normalized using such quantity. For slender (anisotropic) objects (\cref{fig:source}\emph{b}), we observe instead that the fiber length $c$ does not appreciably change the position of the peak of $F_\mathrm{fs}$, it rather appears to be controlled by the fiber diameter $d$. Differently from spheres, here the fluid-solid contribution shows a wider distribution, therefore suggesting a quantitative role of the fiber length as well, as previously suggested. For both objects, the mass fraction does not appear to control the wavenumber associated with the maximum forcing but only the strength of the backreaction. Remarkably, the same holds also when varying the number of objects $N$.

\subsection{Phenomenological interpretation}

A simple and effective interpretation of our results can be proposed by considering the characteristic Reynolds number experienced by the particles, i.e., $\Rey_\ell = u' \ell/\nu$, in order to argue the main hydrodynamic effect caused by the solid objects and discern peculiar differences between bluff and slender objects. For the sake of simplicity, we consider the root mean square of the fluid velocity fluctuations $u'$ (accounting for its variation due to the effective backreaction) and the sphere diameter $D$ or the fiber diameter $d$ as the reference lengthscale $\ell$. For the spherical particles such choice is natural, whilst for fibers it comes from that previously observed for the energy-transfer balance (\cref{fig:source}\emph{b}). When computing the characteristic Reynolds number, we typically find that for spheres $\Rey_D \sim \mathcal{O}(10^3)$, whereas for fibers $\Rey_d \sim \mathcal{O}(10^1)$.  These estimates suggest that bluff and slender objects experience qualitatively different hydrodynamic regimes, one dominated by inertial and the other by viscous forces, respectively. In particular, spheres are subject to a large-Reynolds-number flow, inducing a turbulent wake on scales comparable with and smaller than $D$, while fibers generate flow structures typical of laminar vortex shedding on scales comparable with $d$, but without any further proliferation of scales due to the dominant viscous dissipation.
Note that here we refer to the range of scales smaller than the characteristic lengthscale associated with the individual particles.
For spheres, the energy of the generated wakes is therefore converted into smaller structures by means of the well-known energy cascading process (controlled by the nonlinear term $\Pi$); for fibers, a similar phenomenology is not possible since the smaller-scale generated flow structures are essentially within the dissipative region.

\begin{figure}
    \centering
    \includegraphics[width=0.5\textwidth]{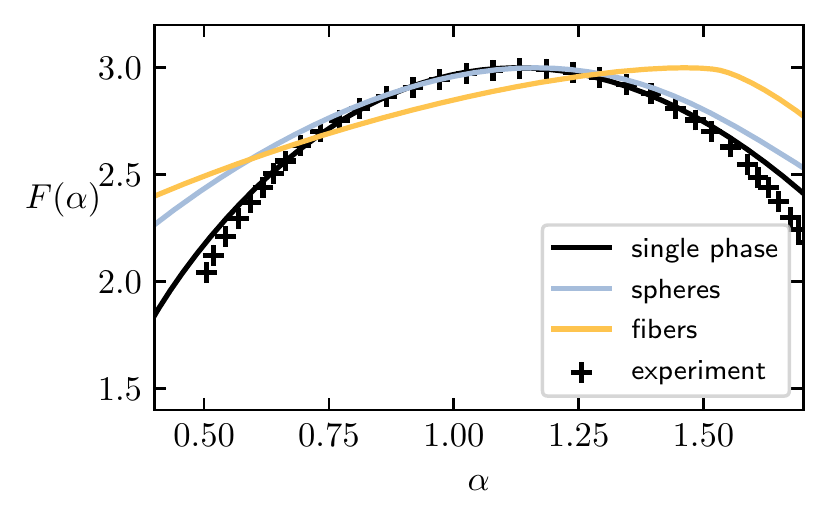}
    \caption{
		Multifractal distribution of the kinetic energy dissipation rate in the single phase flow (black), flow with spheres of mass fraction $M=1$ (blue), flow with fibers of mass fraction $M=1$ (orange), and the single phase experimental measure from~\citet{meneveau} (black crosses). 
		}
    \label{fig:fractal}
\end{figure}

\subsection{On the intermittency of the modulated turbulence}

Strong spatial and/or temporal fluctuations in the energy flux are the source of intermittency in turbulent flows. 
Due to the different nature of the flux in the two configurations, it is natural to wonder how intermittency is altered.
A comprehensive approach to study this is to compute the multifractal spectrum of the energy dissipation rate~\citep{meneveau}, which we report in \cref{fig:fractal}. For spheres, we find that $F(\alpha)$ is substantially similar to the single phase case with only minor differences. On the other hand, for fibers we have a remarkably qualitative difference in the spectrum. This further supports the idea of a standard energy cascade in particle-laden flows with finite-size spherical particles, \added{whilst it is not the case for finite-size fibers}.

\section{Conclusions}
\label{sec:conc}

By means of unprecedented high-Reynolds-number multiphase DNS, we have investigated particle-laden turbulent flows considering solid objects of finite size, i.e. well within the inertial range of scales, with the goal of understanding how the geometrical features of the immersed objects impact on the basic mechanisms of turbulence modulation. Specifically, we have focused on two representative classes of suspensions, i.e. bluff (isotropic) spheres vs slender (anisotropic) fibers, as a benchmark for highlighting the effect of particle anisotropy.

As a common feature, we found that the presence of the dispersed phase induces a similar decrease of the turbulent kinetic energy and micro-scale Reynolds number for increasing mass fraction. At the same time, we unravelled the intrinsic differences in the resulting scale-by-scale energy distribution. For both kinds of dispersed objects we have shown that the representative lengthscale at which the fluid-solid coupling is dominant is associated with the (sphere or fiber) diameter. For finite-size spherical objects, however, the backreaction due to the dispersed phase is always confined to relatively large scales with a negligible alteration of the higher-wavenumber inertial and viscous subrange. Finite-size fibers, on the other hand, transfer energy up to the smallest scales with a consequent modification of the full energy spectrum and the emergence of a modified energy cascade. \added{Note that, while confirming the same phenomenology, these results substantially enrich those recently reported at lower Reynolds number~\citep{olivieri2022fully}, in particular, concerning the evaluation of the scaling exponent in the modulated intermediate range of the energy spectrum. Also, the high-Reynolds-number configuration and the consequent scale separation clarified the different nature of the dominant energy flux in fiber laden flows.}

A simple phenomenological description for this complex problem is that the immersed objects subtract the energy from the flow by means of hydrodynamic drag and then reinject it by their wakes. For spheres, this happens fully within the inertial subrange and therefore results in a turbulent wake that still contributes to the classical energy cascade. For fibers, the transfer involves significantly smaller scales where viscosity eventually dominates, providing to the latter additional energy with little contribution of the nonlinear terms due to the low local Reynolds number.

In conclusion, we underline that these results are unique for finite-size objects and remarkably different from what previously observed for small particles (i.e., whose size is smaller than Kolmogorov's dissipative lengthscale). Our findings have primary relevance for advancing the fundamental understanding of particle-laden turbulence and its numerous related applications (e.g., slurry flows, combustion, papermaking and other industrial processes).

\backsection[Acknowledgements]{
The authors acknowledge the computer time provided by the Scientific Computing section of Research Support Division at OIST and the computational resources of the supercomputer Fugaku provided by RIKEN through the HPCI System Research Project (Project IDs: hp210229 and hp210269).
}

\backsection[Funding]{
The research was supported by the Okinawa Institute of Science and Technology Graduate University (OIST) with subsidy funding from the Cabinet Office, Government of Japan.
}

\backsection[Declaration of interests]{The authors report no conflict of interest.}

\backsection[Data availability statement]{The data that support the findings of this study are available from the corresponding authors upon reasonable request.}

\backsection[Author ORCIDs]{\\
S. Olivieri, \href{https://orcid.org/0000-0002-7795-6620}{https://orcid.org/0000-0002-7795-6620};\\
I. Cannon, \href{https://orcid.org/0000-0002-1676-9338}{https://orcid.org/0000-0002-1676-9338};\\
M. E. Rosti, \href{https://orcid.org/0000-0002-9004-2292}{https://orcid.org/0000-0002-9004-2292}.}

\bibliographystyle{jfm}
\bibliography{references,TriperiodicParticlesBubbles}

\begin{thebibliography}{36}
\expandafter\ifx\csname natexlab\endcsname\relax\def\natexlab#1{#1}\fi
\def\au#1{#1} \def\ed#1{#1} \def\yr#1{#1}\def\at#1{#1}\def\jt#1{\textit{#1}}
  \def\bt#1{#1}\def\bvol#1{\textbf{#1}} \def\vol#1{#1} \def\pg#1{#1}
  \def\publ#1{#1}\def\arxiv#1{#1}\def\org#1{#1}\def\st#1{\textit{#1}}

\bibitem[Alexakis \& Biferale(2018)]{alexakis2018cascades}
{\sc \au{Alexakis, A.} \& \au{Biferale, L.}} \yr{2018}  \at{Cascades and
  transitions in turbulent flows}.  \jt{Phys. Rep.}  \bvol{767},  \pg{1--101}.

\bibitem[Andersson {\em et~al.\/}(2012)Andersson, Zhao \&
  Barri]{andersson2012torque}
{\sc \au{Andersson, H.~I.}, \au{Zhao, L.} \& \au{Barri, M.}} \yr{2012}
  \at{Torque-coupling and particle--turbulence interactions}.  \jt{J. Fluid
  Mech.}  \bvol{696},  \pg{319--329}.

\bibitem[Ardekani {\em et~al.\/}(2019)Ardekani, Rosti \&
  Brandt]{ardekani2019turbulent}
{\sc \au{Ardekani, M.~N.}, \au{Rosti, M.~E.} \& \au{Brandt, L.}} \yr{2019}
  \at{Turbulent flow of finite-size spherical particles in channels with
  viscous hyper-elastic walls}.  \jt{J. Fluid Mech.}  \bvol{873},
  \pg{410--440}.

\bibitem[Balachandar \& Eaton(2010)]{balachandar2010turbulent}
{\sc \au{Balachandar, S.} \& \au{Eaton, J.~K.}} \yr{2010}  \at{Turbulent
  dispersed multiphase flow}.  \jt{Annu. Rev. Fluid Mech.}  \bvol{42},
  \pg{111--133}.

\bibitem[Bounoua {\em et~al.\/}(2018)Bounoua, Bouchet \&
  Verhille]{bounoua2018tumbling}
{\sc \au{Bounoua, S.}, \au{Bouchet, G.} \& \au{Verhille, G.}} \yr{2018}
  \at{Tumbling of inertial fibers in turbulence}.  \jt{Phys. Rev. Lett.}
  \bvol{121}~(12),  \pg{124502}.

\bibitem[Brandt \& Coletti(2021)]{brandt2021particle}
{\sc \au{Brandt, L.} \& \au{Coletti, F.}} \yr{2021}  \at{Particle-laden
  turbulence: Progress and perspectives}.  \jt{Annu. Rev. Fluid Mech.}
  \bvol{54},  \pg{159--189}.

\bibitem[Breard {\em et~al.\/}(2016)Breard, Lube, Jones, Dufek, Cronin,
  Valentine \& Moebis]{breard2016coupling}
{\sc \au{Breard, E. C.~P.}, \au{Lube, G.}, \au{Jones, J.~R.}, \au{Dufek, J.},
  \au{Cronin, S.~J.}, \au{Valentine, G.~A.} \& \au{Moebis, A.}} \yr{2016}
  \at{Coupling of turbulent and non-turbulent flow regimes within pyroclastic
  density currents}.  \jt{Nat. Geosci.}  \bvol{9}~(10),  \pg{767--771}.

\bibitem[Brizzolara {\em et~al.\/}(2021)Brizzolara, Rosti, Olivieri, Brandt,
  Holzner \& Mazzino]{brizzolara2021fiber}
{\sc \au{Brizzolara, S.}, \au{Rosti, M.~E.}, \au{Olivieri, S.}, \au{Brandt,
  L.}, \au{Holzner, M.} \& \au{Mazzino, A.}} \yr{2021}  \at{Fiber tracking
  velocimetry for two-point statistics of turbulence}.  \jt{Phys. Rev. X}
  \bvol{11}~(3),  \pg{031060}.

\bibitem[Cannon {\em et~al.\/}(2021)Cannon, Izbassarov, Tammisola, Brandt \&
  Rosti]{cannon2021effect}
{\sc \au{Cannon, I.}, \au{Izbassarov, D.}, \au{Tammisola, O.}, \au{Brandt, L.}
  \& \au{Rosti, M.~E.}} \yr{2021}  \at{The effect of droplet coalescence on
  drag in turbulent channel flows}.  \jt{Phys. Fluids}  \bvol{33}~(8),
  \pg{085112}.

\bibitem[Capecelatro {\em et~al.\/}(2018)Capecelatro, Desjardins \&
  Fox]{capecelatro2018transition}
{\sc \au{Capecelatro, J.}, \au{Desjardins, O.} \& \au{Fox, R.~O.}} \yr{2018}
  \at{On the transition between turbulence regimes in particle-laden channel
  flows}.  \jt{J. Fluid Mech.}  \bvol{845},  \pg{499--519}.

\bibitem[Cavaiola {\em et~al.\/}(2020)Cavaiola, Olivieri \&
  Mazzino]{cavaiola2019assembly}
{\sc \au{Cavaiola, M.}, \au{Olivieri, S.} \& \au{Mazzino, A.}} \yr{2020}
  \at{The assembly of freely moving rigid fibres measures the flow velocity
  gradient tensor}.  \jt{J. Fluid Mech.}  \bvol{894},  \pg{A25}.

\bibitem[De~Lillo {\em et~al.\/}(2014)De~Lillo, Cencini, Durham, Barry,
  Stocker, Climent \& Boffetta]{delillo2014turbulent}
{\sc \au{De~Lillo, F.}, \au{Cencini, M.}, \au{Durham, W.~M.}, \au{Barry, M.},
  \au{Stocker, R.}, \au{Climent, E.} \& \au{Boffetta, G.}} \yr{2014}
  \at{Turbulent fluid acceleration generates clusters of gyrotactic
  microorganisms}.  \jt{Phys. Rev. Lett.}  \bvol{112}~(4),  \pg{044502}.

\bibitem[Dodd \& Ferrante(2016)]{dodd2016interaction}
{\sc \au{Dodd, M.~S.} \& \au{Ferrante, A.}} \yr{2016}  \at{On the interaction
  of {Taylor} length scale size droplets and isotropic turbulence}.  \jt{J.
  Fluid Mech.}  \bvol{806},  \pg{356--412}.

\bibitem[Falkinhoff {\em et~al.\/}(2020)Falkinhoff, Obligado, Bourgoin \&
  Mininni]{falkinhoff2020preferential}
{\sc \au{Falkinhoff, F.}, \au{Obligado, M.}, \au{Bourgoin, M.} \& \au{Mininni,
  P.~D.}} \yr{2020}  \at{Preferential concentration of free-falling heavy
  particles in turbulence}.  \jt{Phys. Rev. Lett.}  \bvol{125}~(6),
  \pg{064504}.

\bibitem[Freund \& Ferrante(2019)]{freund2019wavelet}
{\sc \au{Freund, A.} \& \au{Ferrante, A.}} \yr{2019}  \at{Wavelet-spectral
  analysis of droplet-laden isotropic turbulence}.  \jt{J. Fluid Mech.}
  \bvol{875},  \pg{914--928}.

\bibitem[Gualtieri {\em et~al.\/}(2013)Gualtieri, Picano, Sardina \&
  Casciola]{gualtieri2013clustering}
{\sc \au{Gualtieri, P.}, \au{Picano, F.}, \au{Sardina, G.} \& \au{Casciola,
  C.~M.}} \yr{2013}  \at{Clustering and turbulence modulation in particle-laden
  shear flows}.  \jt{J. Fluid Mech.}  \bvol{715},  \pg{134--162}.

\bibitem[Hori {\em et~al.\/}(2022)Hori, Rosti \& Takagi]{hori2022eulerian}
{\sc \au{Hori, N.}, \au{Rosti, M.~E.} \& \au{Takagi, S.}} \yr{2022}  \at{An
  {Eulerian-based} immersed boundary method for particle suspensions with
  implicit lubrication model}.  \jt{Comput. Fluids}  \pg{p. 105278}.

\bibitem[Huang {\em et~al.\/}(2007)Huang, Shin \& Sung]{huang_shin_sung_2007a}
{\sc \au{Huang, W.-X.}, \au{Shin, S.~J.} \& \au{Sung, H.~J.}} \yr{2007}
  \at{Simulation of flexible filaments in a uniform flow by the immersed
  boundary method}.  \jt{J. Comput. Phys.}  \bvol{226}~(2),  \pg{2206 -- 2228}.

\bibitem[Hwang \& Eaton(2006)]{hwang2006homogeneous}
{\sc \au{Hwang, W.} \& \au{Eaton, J.~K.}} \yr{2006}  \at{Homogeneous and
  isotropic turbulence modulation by small heavy {($St\sim 50$)} particles}.
  \jt{J. Fluid Mech.}  \bvol{564},  \pg{361--393}.

\bibitem[Lucci {\em et~al.\/}(2010)Lucci, Ferrante \&
  Elghobashi]{lucci2010modulation}
{\sc \au{Lucci, F.}, \au{Ferrante, A.} \& \au{Elghobashi, S.}} \yr{2010}
  \at{Modulation of isotropic turbulence by particles of {Taylor} length-scale
  size}.  \jt{J. Fluid Mech.}  \bvol{650},  \pg{5--55}.

\bibitem[Lucci {\em et~al.\/}(2011)Lucci, Ferrante \&
  Elghobashi]{lucci2011stokes}
{\sc \au{Lucci, F.}, \au{Ferrante, A.} \& \au{Elghobashi, S.}} \yr{2011}
  \at{Is {Stokes} number an appropriate indicator for turbulence modulation by
  particles of {Taylor}-length-scale size?}  \jt{Phys. Fluids}  \bvol{23}~(2),
  \pg{025101}.

\bibitem[Maxey(2017)]{maxey2017simulation}
{\sc \au{Maxey, M.}} \yr{2017}  \at{Simulation methods for particulate flows
  and concentrated suspensions}.  \jt{Annu. Rev. Fluid Mech.}  \bvol{49},
  \pg{171--193}.

\bibitem[Olivieri {\em et~al.\/}(2020{\natexlab{{\em a\/}}})Olivieri, Akoush,
  Brandt, Rosti \& Mazzino]{olivieri2020turbulence}
{\sc \au{Olivieri, S.}, \au{Akoush, A.}, \au{Brandt, L.}, \au{Rosti, M.~E.} \&
  \au{Mazzino, A.}} \yr{2020{\natexlab{{\em a\/}}}}  \at{Turbulence in a
  network of rigid fibers}.  \jt{Phys. Rev. Fluids}  \bvol{5}~(7),
  \pg{074502}.

\bibitem[Olivieri {\em et~al.\/}(2020{\natexlab{{\em b\/}}})Olivieri, Brandt,
  Rosti \& Mazzino]{olivieri2020dispersed}
{\sc \au{Olivieri, S.}, \au{Brandt, L.}, \au{Rosti, M.~E.} \& \au{Mazzino, A.}}
  \yr{2020{\natexlab{{\em b\/}}}}  \at{Dispersed fibers change the classical
  energy budget of turbulence via nonlocal transfer}.  \jt{Phys. Rev. Lett.}
  \bvol{125}~(11),  \pg{114501}.

\bibitem[Olivieri {\em et~al.\/}(2021)Olivieri, Mazzino \&
  Rosti]{olivieri2021universal}
{\sc \au{Olivieri, S.}, \au{Mazzino, A.} \& \au{Rosti, M.~E.}} \yr{2021}
  \at{Universal flapping states of elastic fibers in modulated turbulence}.
  \jt{Phys. Fluids}  \bvol{33}~(7),  \pg{071704}.

\bibitem[Olivieri {\em et~al.\/}(2022)Olivieri, Mazzino \&
  Rosti]{olivieri2022fully}
{\sc \au{Olivieri, S.}, \au{Mazzino, A.} \& \au{Rosti, M.~E.}} \yr{2022}
  \at{On the fully coupled dynamics of flexible fibres dispersed in modulated
  turbulence}.  \jt{J. Fluid Mech.}  \bvol{946},  \pg{A34}.

\bibitem[Podvigina \& Pouquet(1994)]{podvigina1994non}
{\sc \au{Podvigina, O.} \& \au{Pouquet, A.}} \yr{1994}  \at{On the non-linear
  stability of the 1: 1: 1 {ABC} flow}.  \jt{Phys. D: Nonlinear Phenomen.}
  \bvol{75}~(4),  \pg{471--508}.

\bibitem[Rosti \& Brandt(2020)]{rosti2020increase}
{\sc \au{Rosti, M.~E.} \& \au{Brandt, L.}} \yr{2020}  \at{Increase of turbulent
  drag by polymers in particle suspensions}.  \jt{Phys. Rev. Fluids}
  \bvol{5}~(4),  \pg{041301}.

\bibitem[Rosti {\em et~al.\/}(2021)Rosti, Cavaiola, Olivieri, Seminara \&
  Mazzino]{rosti2021turbulence}
{\sc \au{Rosti, M.~E.}, \au{Cavaiola, M.}, \au{Olivieri, S.}, \au{Seminara, A.}
  \& \au{Mazzino, A.}} \yr{2021}  \at{Turbulence role in the fate of
  virus-containing droplets in violent expiratory events}.  \jt{Phys. Rev.
  Research}  \bvol{3}~(1),  \pg{013091}.

\bibitem[Rosti {\em et~al.\/}(2019)Rosti, Ge, Jain, Dodd \&
  Brandt]{rosti2019droplets}
{\sc \au{Rosti, M.~E.}, \au{Ge, Z.}, \au{Jain, S.~S.}, \au{Dodd, M.~S.} \&
  \au{Brandt, L.}} \yr{2019}  \at{Droplets in homogeneous shear turbulence}.
  \jt{J. Fluid Mech.}  \bvol{876},  \pg{962--984}.

\bibitem[Rosti {\em et~al.\/}(2020)Rosti, Olivieri, Cavaiola, Seminara \&
  Mazzino]{rosti2020fluid}
{\sc \au{Rosti, M.~E.}, \au{Olivieri, S.}, \au{Cavaiola, M.}, \au{Seminara, A.}
  \& \au{Mazzino, A.}} \yr{2020}  \at{Fluid dynamics of {COVID-19} airborne
  infection suggests urgent data for a scientific design of social distancing}.
   \jt{Sci. Rep.}  \bvol{10}~(1),  \pg{1--9}.

\bibitem[Sengupta {\em et~al.\/}(2017)Sengupta, Carrara \&
  Stocker]{sengupta2017phytoplankton}
{\sc \au{Sengupta, A.}, \au{Carrara, F.} \& \au{Stocker, R.}} \yr{2017}
  \at{Phytoplankton can actively diversify their migration strategy in response
  to turbulent cues}.  \jt{Nature}  \bvol{543}~(7646),  \pg{555--558}.

\bibitem[Sreenivasan \& Meneveau(1988)]{meneveau}
{\sc \au{Sreenivasan, K.~R.} \& \au{Meneveau, C.}} \yr{1988}  \at{Singularities
  of the equations of fluid motion}.  \jt{Phys. Rev. A}  \bvol{38}~(12),
  \pg{6287--6295}.

\bibitem[Uhlmann \& Chouippe(2017)]{uhlmann_clustering_2017}
{\sc \au{Uhlmann, M.} \& \au{Chouippe, A.}} \yr{2017}  \at{Clustering and
  preferential concentration of finite-size particles in forced
  homogeneous-isotropic turbulence}.  \jt{Journal of Fluid Mechanics}
  \bvol{812},  \pg{991--1023}.

\bibitem[Wang {\em et~al.\/}(2022)Wang, Yi, Jiang \& Sun]{wang2022finite}
{\sc \au{Wang, C.}, \au{Yi, L.}, \au{Jiang, L.} \& \au{Sun, C.}} \yr{2022}
  \at{How do the finite-size particles modify the drag in {Taylor--Couette}
  turbulent flow}.  \jt{J. Fluid Mech.}  \bvol{937}.

\bibitem[Yousefi {\em et~al.\/}(2020)Yousefi, Ardekani \&
  Brandt]{yousefi_modulation_2020}
{\sc \au{Yousefi, A.}, \au{Ardekani, M.~N.} \& \au{Brandt, L.}} \yr{2020}
  \at{Modulation of turbulence by finite-size particles in statistically
  steady-state homogeneous shear turbulence}.  \jt{Journal of Fluid Mechanics}
  \bvol{899},  \pg{A19}.

\end{thebibliography}

\end{document}